\documentstyle[epsf,feynmf]{elsart}
\newcommand{\nn}{\nonumber}
\newcommand{\be}{\begin{equation}}
\newcommand{\ee}{\end{equation}}
\newcommand{\bea}{\begin{eqnarray}}
\newcommand{\eea}{\end{eqnarray}}

\def\siml{{\ \lower-1.2pt\vbox{\hbox{\rlap{$<$}\lower6pt\vbox{\hbox{$\sim$}}}}\ }} 

\begin{document}
\begin{frontmatter}
\begin{flushright}
\tt{UB-ECM-PF 00/07\\ UCSD/PTH 00-09}
\end{flushright}
\vskip 1truecm
\title{Light Fermion Finite Mass Effects in Non-relativistic Bound States}
\author {Dolors Eiras$^1$ and Joan Soto$^{1,2}$}
\address{$^1$ Dept. d'Estructura i Constituents de la Mat\`eria and IFAE, 
     U. Barcelona \\ Diagonal 647, E-08028 Barcelona, Catalonia, Spain}
\address{$^2$ Department of Physics, University of California at San Diego,\\
     9500 Gilman Drive, La Jolla, CA 92093}
\begin{abstract}
We present analytic expressions for the vacuum polarization effects due to a light fermion with finite mass  
in the binding energy and in the wave function at the origin of QED and (weak coupling) QCD non-relativistic 
bound states. Applications to exotic atoms,
 $\Upsilon (1s)$ and $t\bar t$ production near threshold are briefly discussed.
\end{abstract}
\vspace{1cm}
{\small PACS numbers: 11.10.St, 13.25.Gv, 36.10-k}
\end{frontmatter}

\newpage        

\pagenumbering{arabic}

\section{Introduction}

It is well known that the vacuum polarization effects due to light fermions produce leading corrections to
the Coulomb-like behavior of non-relativistic bound states both in QED \cite{DiGiacomo} and in QCD 
\cite{Billore}. If a light
 particle has a mass ($m_l$) of the order of the inverse Bohr radius, it can neither be approximated by a massless
particle  nor by a very heavy one. Hence any observable related to the bound state depends non-trivially on
 $m_l$. For a given bound state the $m_l$ dependence in physical observables is usually 
calculated
 numerically \cite{DiGiacomo,Borie,Savely3,Labelle,Pachucki,Kinoshita,Melles} and only a few analytical results are available
 \cite{pionium,Pustovalov,Savely1,Savely2}\footnote{We learnt about the three last references after completing our calculations.}. In this letter we present further analytical formulas 
with the exact light fermion mass dependence  for the leading corrections to the energy shift for arbitrary 
quantum numbers $(n,l)$ and to the wave function at the origin for the ground state.    

These formulas may be useful for quite a few physical systems of current interest. 
Any QED bound state built out
of particles heavier than the electron may require them. For instance di-muonium, muonic hydrogen, pionium, 
pionic hydrogen and other 
simple hadronic atoms where the electron mass ($m_e$) is such that $m_e\sim \mu \alpha /n$, $\mu$ and $n$ being 
the reduced mass and the principle quantum number respectively. It is worth mentioning that simple hadronic atoms
are experiencing a renewed interest because they may allow to extract important information on the 
QCD scattering lengths for several isospin channels \cite{L}.
In particular the measurement of the decay width of pionium \cite{Dirac}
 at the $10\%$ level will allow to extract a combination of scattering lengths 
with sufficient accuracy as to 
discern between the large and the small quark condensate scenario of QCD, namely
 between Chiral Perturbation Theory \cite{GL} and Generalized Chiral Perturbation Theory \cite{Stern}.    
In fact, the leading corrections to the pionium decay width have been recently calculated in a systematic way
using non-relativistic effective field theory techniques \cite{pionium,Bern} (see \cite{relativistic},
\cite{Labelle,RavndalHolstein} and \cite{nuclear} for earlier 
relativistic, non-relativistic, and quantum mechanical calculations respectively). Fully analytic results have been 
obtained except for the contribution of the electron mass to the vacuum polarization where only a numerical 
result is available \cite{Labelle}. We shall fill this gap here and present the only remaining piece to have 
the full leading corrections to the pionium decay width in an analytic form. 

For QCD non-relativistic bound states, $\Upsilon (1s)$ seems to be the only one amenable to a weak coupling analysis
\cite{VoloshinLeutwylerPineda}. Using the results of \cite{pionium}, analytic expressions for the binding 
energy shift due to the finite charm mass have been recently presented in \cite{HoangManohar}\footnote{The extraction of the bottom $\overline{\rm MS}$ mass from the $\Upsilon (1s)$ mass also requires the bottom pole mass dependence on $m_c$ \cite{Gray}.}. We present here 
analytic results for these effects in the wave function at the origin.

An important non-relativistic weak coupling QCD system for the Next Linear Collider physics is the top 
quark-antiquark pair near threshold. The production cross-section has already  
been calculated at NNLO \cite{top}. However, the calculations are done assuming the mass of the bottom and charm
quarks zero. Our results may allow to include the leading effects of these masses.

The fact that our results can be applied to such a variety of a priory quite different physical systems can be
easily understood in terms of modern effective field theories. Non-relativistic bound states have at least three 
dynamical scales: the hard scale (mass of the particles forming the bound state), the soft scale (typical relative
momentum in the bound state) and the ultrasoft scale (typical binding energy in the bound state). Upon integrating
out the hard scale local non-relativistic effective theories arise. These are NRQED for QED, NRQCD for 
QCD \cite{CL}, and NR$\chi$L for the Chiral Lagrangian\footnote{We apologize to the authors of ref. \cite{Labelle}
for slightly modifying their previously given name, namely {\it Non Relativistic Chiral Perturbation Theory}.
The reason is that the calculation in the non-relativistic regime cannot be 
organized according to the chiral counting any longer \cite{pionium}.}
\cite{Zuoz}.
Upon integrating out the soft scale
(see \cite{Pos,pNRQCD} for the precise statements) effective theories which are local in time but non-local in 
space arise. The non-local terms in space are nothing but the usual quantum-mechanical potentials and only
 ultrasoft degrees of freedom\footnote{In the language of the threshold expansions in QCD \cite{BSS} these 
correspond to ultrasoft gluons and potential quarks.} 
are left dynamical. The corresponding non-relativistic effective theories have been 
named pNRQED, pNRQCD and pNR$\chi$L for QED, QCD and the Chiral Lagrangian respectively \cite{Zuoz,Mont}, 
(the ``p'' stands for potential).
Since the leading (mass independent) coupling of the photon field
to the non-relativistic charged particles as well as the one of the gluon field to the non-relativistic quarks 
in the NR theories is universal, it produces the same potential in the pNR theories, and hence they all can be 
discussed at once. If there is a light (relativistic) charged particle in QED or a light (relativistic) 
quark in QCD  whose mass is of the order of
 the soft scale, it must be integrated out keeping the mass dependence exact, which produces a  
light fermion mass dependent correction to the static potential. 

\begin{fmffile}{fee1}
 \begin{center}

 \begin{fmfgraph*}(80,80)
\fmfleft{i1,i2}\fmfright{o1,o2}
 \fmf{fermion,label=$h^{\prime}$,label.side=left}{i1,v1}
 \fmf{fermion,label=$\bar{h}$}{i2,v2} 
\fmf{dashes}{v2,v3}
\fmf{dashes}{v4,v1}
\fmf{plain,left,label=$l$,label.side=left}{v3,v4}
\fmf{plain,right,label=$\bar{l}$,label.side=right}{v3,v4}
\fmf{fermion,label=$h^{\prime}$,label.side=left}{v1,o1} 
\fmf{fermion, label=$\bar{h}$,label.side=right}{v2,o2}
\end{fmfgraph*}
\begin{fmfgraph*}(90,90)
\fmfleft{a}
\fmfright{b}
\fmftop{tp}
\fmfbottom{bt}
\fmf{phantom}{a,x,b}
\fmf{phantom}{tp,c1,c2,bt}
\fmfv{label= $ = $}{x}
\end{fmfgraph*} 
\begin{fmfgraph*}(90,90)
\fmfleft{i}
\fmfright{o}
\fmftop{t}
\fmfbottom{b}
\fmf{dbl_plain}{i,v,o}
\fmf{phantom}{t,v1,v2,b}
\fmfv{label=$V_{vpc}$}{v1}
\fmfv{decor.shape=triangle, decor.filled=shaded, decor.size=50}{v}
 \end{fmfgraph*} 
 \end{center}
\end{fmffile} 

\begin{center}
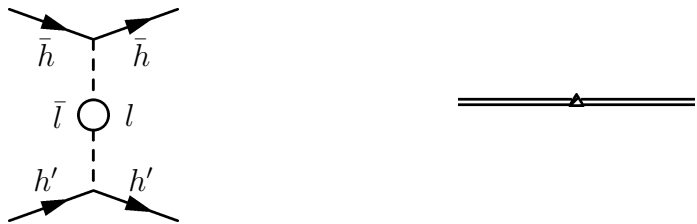
{\footnotesize Fig.1: Matching between the non-relativistic theory and the potential one.} 
\end{center}
When matching the NR theories to the pNR theories only the diagram of Fig. 1 gives rise to a potential which 
contributes to the leading effect. For QED (on-shell scheme) it reads,
\be  
V_{vpc}(\vert {\bf x} \vert) \, = \, -\frac{\alpha}{\pi}\frac{\alpha}{\vert {\bf x }\vert } \int_{\small 0}^{\small 1} \, dv \, \frac{v^2 \left ( 1-\frac{v^2}{3} \right )}{(1-v^2)} e^{-\frac{m_l \vert {\bf x }\vert }{\sqrt{1-v^2}}}
\ee
and for QCD ($\overline{\rm MS}$): 
\bea
V_{vpc}(\vert {\bf x} \vert) \, &=& \, -\frac{C_{\small F} T_{\small F} \alpha_s}{\pi}\frac{\alpha_s}{\vert {\bf x }\vert } \left \{ \int_{\small 0}^{\small 1} \, dv \, \frac{v^2 \left ( 1-\frac{v^2}{3} \right )}{(1-v^2)} e^{-\frac{m_l \vert {\bf x }\vert }{\sqrt{1-v^2}}}+ \frac{1}{3} \log
 \left ( \frac{m_l^2}{\nu^2} \right ) \right \} \nn \\
C_{\small F} \,&=& \, \frac{N_{\small c}^2 -1}{2N_{\small c}}=\frac{4}{3} \quad , \quad 
T_{\small F} \, = \, \frac{1}{2}
\eea
If $N_f$ is the number of flavors lighter that $m_l$, the 
$\alpha_s (\nu)$ above runs with $N_f+1$
flavors. Notice that the difference between the QED and the QCD case is, apart from the trivial color factors
$\alpha /\vert {\bf x}\vert\rightarrow C_F\alpha_{s} /\vert {\bf x}\vert$ and $\alpha /\pi\rightarrow T_{F}\alpha_{s}/\pi$,
a term which 
can be absorbed in a redefinition of the Coulomb potential \cite{TY}. Hence for the actual calculation we shall
 only deal with (1) and use these facts to extend our results to the QCD realm.

\section{Energy Shift}

For the energy shift we obtain ($\xi \, := \, \frac{n m_l}{\mu \alpha}$):
\bea
\delta E_{nl} (\xi ) \,& =& \, -\frac{2\alpha}{3\pi} E_n \left \{ \frac{5}{3} - \frac{3\pi}{2}n \xi + \left ( n(2n+1)+(n+l)(n-l-1) \right ) \xi^2 - 
\right. \nonumber \\ 
&& \left.- \pi n \left ( \frac{1}{3} (n+1)(2n+1) + (n+l)(n-l-1) \right ) \xi^3 - \right. \nn \\
&& \left. - 
\frac{1}{(2n-1)!} \sum_{k=0}^{n-l-1} \, \left ( \begin{array}{c} n-l-1 \\ k \end{array} \right ) \left ( \begin{array}{c}  n+l \\  2l+1+k \end{array} \right ) \xi^{2(n-l-1-k)} \right. \nn \\
&& \left.\frac{d^{2n-1}}{d\xi^{2n-1}} \, \left [ \xi^{2(k+l)+1} (2-\xi^2-\xi^4) F_1(\xi) \right ] \right \} 
\eea
\bea
F_1 (\xi )  \,   := \, \left \{ \begin{array}{cl}
\frac{1}{\sqrt{\xi^2 -1}} \arccos \frac{1}{\xi} & \mbox{if } \, \, \xi > 1, \nn \\ 1 & \mbox{if } \, \,\xi = 1, \nn \\  \frac{1}{\sqrt{1-\xi^2}} \log \left [ \frac{1+\sqrt{1-\xi^2}}{\xi} \right ] & \mbox{if } \, \, \xi<1.
\end{array} \right.
\eea
where $E_{n}=-\mu \alpha^2/2n^2$ is the Coulomb energy.
For  $\xi$ large, namely $m_l >> \mu \alpha /n$, it reduces to
\bea
\delta E_{nl} (\xi \,  \rightarrow \, \infty ) \,  \rightarrow  \, \frac{2\alpha}{3\pi} \frac{E_n}{\xi^{2l+2}} \left \{ \frac{(n+l)!}{(n-l-1)!(2l+1)!} \frac{(2l)!!}{(2l+1)!!} \right. 
\nn \\  \left. \left ( 2- \frac{(2l+2)}{(2l+3)} - \frac{(2l+2)(2l+4)}{(2l+3)(2l+5)} + O \left ( {1\over \xi}\right ) \right ) \right \},
\eea
whereas for $\xi$ small, namely $m_l << \mu \alpha /n$, we obtain
\bea
\delta E_{nl} (\xi \,  \rightarrow \, 0) \,  \rightarrow  \, -\frac{2\alpha}{3\pi} E_n \left \{ \frac{5}{3}+ 2 \left ( \psi(n+l+1)-\psi(1) \right )- 2\log \frac{2}{\xi} -\right.  \nn \\
\left.  -\frac{3\pi}{2}n\xi + \frac{3}{2} \left ( n(2n+1)+(n+l)(n-l-1) \right ) \xi^2 - \right. \nn \\
\left. -\pi n \left ( \frac{1}{3} (n+1)(2n+1)+(n+l)(n-l-1) \right )\xi^3 +O(\xi^4) \right \} 
\eea
where we have used (13). The key steps to obtain (3) are given in the Appendix B. 
We have done the following checks. For the 1S state (3) reduces to the formula (5.3) of ref. \cite{pionium}. 
The energy shifts for the 1S,
 2S, 2P, 3S, 3P and 3D states agree with the early analytical formulas of 
ref. \cite{Pustovalov}.
For $\xi$ large, we reproduce the well-known
positronium like limit for $l=0$ (to be precise we agree with the correction to the energy obtained using formula (2.8) of
\cite{Pos}). We also agree for $l=1$ with formula (32) of ref. \cite{Savely3}. 
For $\xi$ small, 
we can compare with known results for massless quarks in QCD. For arbitrary $n$ and $l$ we agree with formula (13) of ref. \cite{TY}.
For $l=n-1$ we agree with $O(\xi^0)$ and $O(\xi^1)$ of formula (14) in ref. \cite{Savely2}\footnote{Taking
 $\epsilon_{n}=0$
 in that reference
 and upon correcting an obvious misprint $\kappa_1\rightarrow \kappa_n$.} but disagree with their $O(\xi^2)$ result
 (the $O(\xi^3)$ is not displayed in \cite{Savely2}). Notice that for $\xi$ large enormous cancellations occur in formula (3) and hence the analytic expansion (4) may prove very useful.

\section{Wave Function at the Origin}

The correction for the wave function at the origin for $n=1$ states reads
\bea
\delta \Psi_{10} ({\bf 0}) \, & = & \, -\frac{\alpha}{\pi} \Psi_{10} ({\bf 0}) \left [ \left \{ \frac{5}{9} - \frac{\pi}{4}\xi + \frac{1}{3}\xi^2- \frac{\pi}{6}\xi^3+\frac{1}{3}(\xi^4+\xi^2-2) F_{\small 1} (\xi) \right \}+ \right. \nn \\
&& \left. \left \{ \frac{11}{18}- \frac{2}{3}\xi^2+\frac{2\pi}{3}\xi^3-\frac{1}{6}(12\xi^4+\xi^2+2)F_{\small 1} (\xi) - \frac{1}{6} \frac{(4\xi^4+\xi^2-2)}{(\xi^2-1)}(1-\xi^2 F_{\small 1}(\xi )) \right \}+ \right. \nn \\
&& \left. + \left \{ \frac{2}{3}+ \frac{\pi}{4}\xi- \frac{1}{9}\xi^2+ \frac{13\pi}{18}\xi^3- \frac{1}{9} (13\xi^4-11\xi^2-11)F_{\small 1}(\xi)- \right. \right. \nn \\
&& \left. \left. -\frac{1}{3}(4\xi^3+3\xi)F_{\small 2}(\xi ) + \frac{1}{3} (4\xi^4+\xi^2 -2) F_{\small 3}(\xi)
+ \frac{1}{3} \left ( 4\xi^2+ \frac{11}{3} \right ) \log \frac{\xi}{2} 
\right \} \right ]  
\eea
where $\Psi_{10} ({\bf x})$ is the Coulomb wave function. The first bracket corresponds to the zero photon exchange and has already been calculated analytically in \cite{Labelle}.
 The second and third brackets correspond to the pole subtraction and multi-photon exchange contributions respectively.
$F_{i}(\xi)$ , $i=2,3$ are defined as follows:
\bea
F_2(\xi) \, &=& \, \int_{\small 0}^{{\small \frac{\pi}{2}}} \, d\theta \, \log \left [{\sin\theta + \xi\over \sin\theta} \right]  \nn \\
F_3(\xi) \, &=& \, \int_{\small 0}^{{\small \frac{\pi}{2}}} \, d\theta \, \frac{1}{\sin\theta + \xi} \log \left [ \frac{\sin\theta + \xi}{\sin\theta } \right ]
\eea
$F_2(\xi)$ and $F_3(\xi)$ can be expressed in terms of Clausen integrals and dilogarithms. We present
 the explicit formulas in Appendix A. 
The key steps in order to obtain (6) are given in Appendix B.   
\vspace{1mm}
\begin{fmffile}{sarg}
\begin{center}
 \begin{fmfgraph*}(90,90)
 \fmfleft{i}\fmfright{o}\fmftop{t1,t2}\fmfbottom{b1,b2}
 \fmf{dbl_plain}{i,v1,v2,o} 
 \fmf{phantom}{t1,v3,v1,v4,b1}
 \fmf{phantom}{t2,v5,v2,v6,b2}
 \fmfv{label=$V_{vpc}$}{v3}
 \fmfv{decor.shape=triangle,decor.filled=shaded,decor.size=50}{v1}
 \fmfv{label=$\delta ({\bf x})$}{v5}
 \fmfv{decor.shape=hexagram,decor.filled=full,decor.size=50}{v2}
 \end{fmfgraph*}
 \end{center}
 \end{fmffile}

\begin{center}
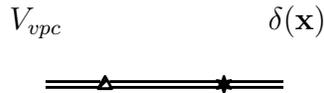
{\footnotesize Fig.2: Diagram rendering the correction to the wave function at the origin. The double line is the Coulomb propagator of the non-relativistic pair and the star a local ($\delta ({\bf x})$) potential.}
\end{center} 
For  $\xi$ large, namely $m_l >> \mu \alpha$, (6) behaves like
\bea
\delta \Psi_{10} ({\bf 0})_{\small \xi \rightarrow \infty} \, \rightarrow \, \frac{\alpha}{\pi} \Psi_{10} ({\bf 0}) \left [ \frac{3\pi}{16\xi} +\frac{107}{225\xi^2}+\frac{4}{15\xi^2}\log \frac{\xi}{2} + O\left ( {1\over \xi^3} \right ) \right ]
\eea
This result must be compatible with the one obtained by integrating out the light fermion first and then calculating 
the electromagnetic potential. In the case $m>> m_l >> \mu\alpha /n$  (for simplicity, we are assuming $h=h^{\prime}$,
 $m$ being the mass of the non-relativistic particles) we expect that a local non-relativistic effective theory is 
obtained after integrating out the energy scale $m_l$ and the associated three momentum scale $\sqrt{m m_l}$ for the 
non-relativistic particle. The leading term in (8) corresponds to the contribution that would be obtained from 
the local
term induced by the diagram in Fig. 3. The logarithm in the subleading term corresponds to the iteration of two 
delta function potentials in quantum mechanics (see formula (5.5) in \cite{pionium}). The second delta function 
is due 
to the contribution to the electromagnetic potential of the dimension six photon operator (see \cite{Lamb})
 which arises after integrating out a heavy particle \cite{Ball}.\\
\vspace{1mm}
\begin{fmffile}{eye}
\begin{center}
\begin{fmfgraph*}(90,90)
\fmfleft{i1,i2}
\fmfright{o1,o2} 
\fmfpen{thick}
\fmf{plain,label=$h^{\prime}$,label.side=right}{i1,v2}
\fmf{plain}{v2,v21,v3,v4,v1,v5,v6,v34,v7}
\fmf{plain,label=$h^{\prime}$,label.side=right}{v7,o1}
\fmf{plain,label=$\bar{h}$,label.side=left}{i2,v8}
\fmf{plain}{v8,v22,v9,v10,v1,v11,v12,v35,v13}
\fmf{plain,label=$\bar{h}$,label.side=left}{v13,o2}
\fmfv{decor.shape=hexagram,decor.filled=full,decor.size=50}{v1}
\fmffreeze
\fmf{phantom}{v2,v14,v15,v16,v8}
\fmfpen{thin}
\fmf{dashes}{v2,v14}
\fmf{dashes}{v16,v8}
\fmf{plain,left}{v14,v16,v14}
\fmf{phantom,left,label=$l$}{v14,v16}
\fmf{phantom,right,label=$\bar{l}$}{v14,v16}
\end{fmfgraph*}
\end{center}
\end{fmffile}

\begin{center}
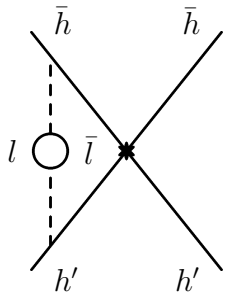
{\footnotesize Fig.3: Vacuum polarization correction to the decay width at leading order in $\frac {1}{\xi}$ when $\xi \rightarrow \infty $. Energies and momenta of order $m_l$ and $\sqrt{m m_l}$ respectively dominate the graph and hence the Coulomb resummation leads to subleading effects. }
\end{center}
For $\xi$ small, namely $m_l << \mu \alpha $, we obtain 
\bea
\delta \Psi_{10}({\bf 0})(\xi \, \rightarrow \, 0) \, \rightarrow \, -\frac{\alpha}{\pi}\Psi_{10}({\bf 0}) \left [ \frac{3}{2}-\frac{\pi^2}{9}-\log \frac{2}{\xi} - \frac{3}{2}\xi^2  + O(\xi^3) \right ]
\eea
We have made the following checks.  For  $\xi$ large and small, the leading term of (8) and (9) agree  with formulas (22)
and (23) of ref. \cite{Savely3} respectively
(the next-to-leading terms are not displayed in \cite{Savely3}). For $\xi$ small
we can also compare with known results for massless quarks in QCD. We agree with the $O(\alpha_s )$ correction 
of formula (69) of ref. \cite{MY}. We have also checked that the formula (6) reproduces the numerical results 
obtained for di-muonium and pionium in refs. \cite{Savely1} and \cite{Labelle} respectively, and
we also agree numerically with the analytical result in terms of a non-trivial integral of ref. \cite{Savely3}.
  
\section{Applications}

\subsection{Exotic Atoms}

We have listed in Table I and Table II the corrections to some energy splittings and to the wave function at the origin 
respectively of simple exotic atoms of current
interest. This purely electromagnetic corrections must be conveniently taken into account if one wants to obtain precise 
information of 
the strong scattering lengths from hadronic atoms.
\begin{center}
{\bf TABLE I}
\vspace{.5cm}
\\
\begin{tabular}{||c|c|c|c||}
\hline
\hline
 & $\frac{\delta E_{21} - \delta E_{10}}{\alpha (E_{2}-E_{1})}$ &  $\frac{\delta E_{31} - \delta E_{10}}{\alpha (E_{3}-E_{1})}$ & $\frac{\delta E_{21} - \delta E_{20}}{\alpha E_{2}}$ \\
\hline
$pK^{-}$ & .44593 & .38629 & -.10453  \\
$p\pi^{-} $ & .18103 & .15388 & -.056337 \\
$p\mu^{-} $ & .13616 & .11548 & -.044443 \\
\hline
\hline
\end{tabular}
\\
\vspace{.5cm}
\footnotesize{Vacuum polarization induced energy splittings for some exotic atoms.}
\normalsize
\\
\end{center}

\begin{center}

{\bf TABLE II}
\\
\vspace{.5cm}
 
\begin{tabular}{||c|c|c|c|c|c||}
\hline
\hline
 
& $\xi=\frac{m_e}{\mu\alpha}$ & $\frac{\delta_{zph} \Psi({\bf 0})} {\alpha \Psi({\bf 0})}$ &
 $\frac{\delta_{ps} \Psi({\bf 0})}{\alpha \Psi({\bf 0})} $ &
 $\frac{\delta_{mph} \Psi({\bf 0})}{\alpha \Psi({\bf 0})} $ &
 $\frac{\delta \Psi({\bf 0})}{\alpha \Psi({\bf 0})} $ 
\\
\hline
\hline
$K^- p$ & .21648 & .34290  & .15454 & .09650 & .59394 \\ 
$K^+ K^-$ & .28369 & .29837 & .12958 & .08785 & .51581 \\ 
$\pi^- p$ & .57635 & .19613 & .07285 & .06166 & .33064 \\ 
$K^+ \pi^- $ & .64357 & .18237 & .06549 & .05741 & .30527 \\
$\mu^- p$ & .73738 & .16627 & .05703 & .05222 & .27552 \\
$\pi^+ \pi^-$ & 1.00344 & .13338 & .04052 & .04099 & .21490 \\ 
$\mu^+ \mu^-$ & 1.32550 & .10793 & .02876 & .03184 & .16853 \\
\hline
\hline
\end{tabular} \\
\vspace{.5cm}
\footnotesize{Vacuum polarization correction to the ground state wave function at the origin of some exotic atoms.}
\end{center}

\subsection{$\Upsilon (1s)$ and $t\bar{t}$}

The current calculations of heavy quarks near threshold assume that the remaining lighter quarks are massless. 
This approximation is difficult to justify {\it a priori} at least in two cases. For the $\Upsilon (1s)$ system the typical
 relative momentum $m_b \alpha_s /2 \sim$ 1.3 GeV. \cite{PY} is of the same order as the charm mass
$m_c \sim$ 1.5 GeV. The effects of 
a finite charm mass in the binding energy have been recently quantified in \cite{HoangManohar}. We give in Table III the size 
of these effects in the wave function at the origin. For the $t\bar{t}$ production near threshold at a relative
 momentum 
$m_t \alpha_s /2 \sim$ 18 GeV. the effects of a finite bottom mass $m_b\sim$ 5 GeV. should be noticeable.
In order to estimate them, we also show in Table III the size of this effect, both for bottom and charm, in the wave function at 
the origin for the would-be-toponium $(1s)$ state. 
\begin{center}
{\bf TABLE III}
\\
\vspace{.5cm}
 
\begin{tabular}{||c|c|c|c||}
\hline
\hline
 &
$\xi = \frac{m_{c}}{C_F \alpha_s \mu}$ & $\xi = \frac{m_{b}}{C_F \alpha_s \mu}$ &  $\frac {\delta \Psi({\bf 0})_{\xi \neq 0}-\delta \Psi({\bf 0})_{\xi=0}}{\alpha_s\Psi({\bf 0})}$\\
\hline
\hline 
$\bar{b}b$ &  1.4 & & .088 \\
$\bar{t}t$ & & .28 & .011 \\
$\bar{t}t$ & .10 & & .0019 \\
\hline
\hline
\end{tabular} \\
\vspace{.5cm}
\footnotesize{Vacuum polarization correction to wave function at the origin in quarkonia. $\overline{\rm MS}$ has been used.}
\end{center}
\normalsize

If the corrections were organized in a series of $\alpha_s / \pi$ multiplied by numbers of order 1,
one may conclude that the leading effects of a finite quark mass are: (i) in the 
$\Upsilon (1s)$ system for charm more important than the next to leading corrections 
\cite{MY}; 
(ii) in the $t\bar t$ system near threshold for bottom (charm) as important as (less important than) the next to leading corrections 
\cite{top}. However, the relativistic corrections do not have the $\pi$ suppression and some radiative corrections are enhanced by factors of $\beta_0$. 
In practise the next to leading corrections are comparable to the leading ones, even for the $t\bar t$ system (see \cite{Beneke} for a discussion).
This makes the actual size of the finite mass effects smaller than the next to leading order corrections in all the cases above.  

{\bf Acknowledgements} 

We are indebted to Antonio Pineda for providing us with useful references, independent checks of various results 
in the literature and a critical reading of the manuscript. 
We also thank the referee for appropiated remarks on the relative size of the corrections in heavy quark systems.
We are supported by  the AEN98-031 (Spain) and the 1998SGR 00026 (Catalonia).
D.E. acknowledges financial support from a MEC FPI fellowship (Spain) and J.S. from the BGP-08 fellowship
(Catalonia) respectively. J.S. thanks A. Manohar and High Energy Physics Group at UCSD for their warm 
hospitality while this work was  written up. 

{\bf Appendix A}

$F_2(x)$ in (7) can be expressed in terms of Clausen integrals. We get
\bea
F_2(x) = \left \{ \begin{array}{cl}
  2Cl_2(\arcsin\, x)-\frac{1}{2}Cl_2(2\arcsin\, x)
& \mbox{if } x < 1 \cr \\
2 \sum_{k=0}^{\infty} {(-1)^{k}\over (2k+1)^2}\sim 1.831932 & \mbox{if } x = 1 
\cr \\
 -i\,Li_2(-x - {\sqrt{x^2-1}}) + 
   i\,Li_2(i\,\left( -x + {\sqrt{x^2-1}} \right) ) - 
 &\\  -i\,Li_2(-x + {\sqrt{x^2-1}}) + 
 \,i\,Li_2(
         -i\,\left( x + {\sqrt{x^2-1}} \right) )+
   
&\\         
+{\pi\over 8} \,\Bigl( i\pi  +  4\log (2x) - 
 4\log (1 + ix - i{\sqrt{x^2-1}}) - 
          4\log (1 + ix + i{\sqrt{x^2-1}}) \Bigr)+   
&\\      + 2\,\Bigl({\pi\over 2}-i\log\left(\sqrt{x+1\over 2}-\sqrt{x-1\over 2}\right)\Bigr)
\,
        \Bigl( -2\,i\,\arctan ({\frac{x-1}{{\sqrt{x^2-1}}}})+  
&\\         + \log (1 + x - {\sqrt{x^2-1}}) - 
          \log (1 + i\,x - i\,{\sqrt{x^2-1}})+  
&\\         + \log (1 + i\,x + i\,{\sqrt{x^2-1}}) - 
          \log (1 + x + {\sqrt{x^2-1}}) \Bigr)  
 & \mbox{if }  x > 1 .
\end{array} \right.
\eea

Recall that the Clausen integral is defined as
\be
Cl_2(x):= -\int_0^{x} \, d\theta \, \log\left ( 2 \sin{\theta\over 2}\right )
=i{{\pi }^2\over 6} - {\frac{i}{4}}\,{x^2} - x\log (ie^{-i{x\over 2}})-    i\,Li_2({e^{i\,x}})
\ee

$F_3 (x)$ in (7) can be expressed in terms of dilogarithms. We get
\bea
F_3 (x) &= & \frac{1}{\sqrt{1-x^2}} \left [ Li_2 \left (- \frac{(1+a+b)}{2b} \right ) - Li_2 \left (- \frac{(1+a-b)}{2b} \right )+ Li_2 \left (- \frac{2b}{(1+a+b)} \right )- \right. \nn \\ &&\left. -Li_2 \left (\frac{2b}{(1+a+b)} \right )- Li_2 \left ( - \frac{(a+b)}{2b} \right )+Li_2 \left (- \frac{(a-b)}{2b} \right ) - Li_2 \left ( - \frac{2b}{(a+b)} \right )+ \right. \nn \\ && \left. +Li_2 \left ( \frac{2b}{(a+b)} \right )+Li_2 \left (\frac{(a+b)}{(1+a+b)} \right ) - Li_2 \left ( \frac{(a-b)}{(1+a-b)} \right ) + \right. \nn \\ && \left. + \log (b x) \, \log \left ( \frac{(1+a+b)}{(1+a-b)} \right ) \right ] \hspace{6.5cm}    \mbox{if} \quad x < 1, \nn
\eea
where $a:=x^{-1}$ and $b:=\frac{\sqrt{1-x^2}}{x}$. 
\bea
F_3 (1)&=&2 -\log 2 \nn \cr \\
F_3 (x)&=& \frac{1}{i\sqrt{x^2 - 1}} \left [ Li_2 \left (- \frac{(1+a+ib)}{2ib} \right ) - Li_2 \left (- \frac{(1+a-ib)}{2ib} \right )+ Li_2 \left (- \frac{2ib}{(1+a+ib)} \right )- \right. \nn \\ &&\left. -Li_2 \left (\frac{2ib}{(1+a+ib)} \right )- Li_2 \left ( - \frac{(a+ib)}{2ib} \right )+Li_2 \left (- \frac{(a-ib)}{2ib} \right ) - Li_2 \left ( - \frac{2ib}{(a+ib)} \right )+ \right. \nn \\ && \left. +Li_2 \left ( \frac{2ib}{(a+ib)} \right )+Li_2 \left (\frac{(a+ib)}{(1+a+ib)} \right ) - Li_2 \left ( \frac{(a-ib)}{(1+a-ib)} \right ) + \right. \nn \\ && \left. + 2\arctan \left (\frac{b}{1+a} \right ) \left ( i\log (b x ) - \frac{\pi}{2} \right ) \right ] \hspace{5cm}    \mbox{if} \quad x > 1, 
\eea
where $a:= x^{-1}$ and $b:=\frac{\sqrt{x^2-1}}{x}$.\\

In order to make contact with the expressions found in the literature for the massless limit of (3) the following 
formula is useful
\be
\psi (2n)-{(n+l)!(n-l-1)!\over (2n-1)!}\sum_{k=0}^{n-l-2}{(2(n-l-1-k)-1)!(2(k+l)+1)!\over
(n-l-k-1)!^2(2l+1+k)! k!}= \psi(n+l+1)
\ee

{\bf Appendix B}
 
We sketch here the main steps which lead to our analytic formulas. For the energy shift we have to calculate ($v=\sqrt{1-x^2}$)
\bea
\delta E_{\small n l} = <nl\vert V_{vpc}\vert nl>&=& \frac{2\alpha E_{\small n}}{3\pi} \frac{(n+l)!}{(n-l-1)!(2l+1)!} \, \xi^{2n-2l-2}\int_{\small 0}^1
\, dx \frac{x^{2l+1}}{(x+\xi)^{\small 2n}} \sqrt{ 1-x^2} \left ( 2+x^2 \right ) \nn \\ && 
F(-(n-l-1),-(n-l-1), 2l+2 ; {x^2\over \xi^2}) .
\eea
Since $l<n$ the hypergeometric function above reduces to a polynomial
\bea
F(-(n-l-1),-(n-l-1), 2l+2 ;z)&=& \sum_{j=0}^{n-l-1}  \frac{(2l+1)!{(n-l-1)!}^2}{{(n-l-1-j)!}^2(j+2l+1)! j!} 
 z^{j}. \quad
\eea
Hence
\bea
\delta E_{\small n l}& = &\frac{2\alpha E_{\small n}}{3\pi}\sum_{j=0}^{n-l-1} \left ( \begin{array}{c} n-l-1 \\ j \end{array} \right ) \left ( \begin{array}{c} n+l \\ n-l-j-1 \end{array} \right )\xi^{2n-2l-2j-2}\nn \\ &&\times \int_{\small 0}^{\small 1} dx \, 
{x^{2l+2j+1}\over (\xi+
x)^{2n}}
 \sqrt{1-
x^2}
\left( 2+x^2\right)
.\quad
\eea
and upon writing
\be
\left ( {1\over \xi +x}\right ) ^{2n}=-{d^{2n-1}\over d\xi^{2n-1}}{1\over \xi +x}
\ee
and making the change $x\rightarrow \sin\theta$ we obtain (3).

For the wave function at the origin we have to calculate:
\be
\delta \Psi_{\small n0}({\bf 0})\Psi_{n0}({\bf 0})=\lim_{E\rightarrow E_n}<n0\vert \delta ({\bf x})\left({1\over E-H}-{\vert n0><n0\vert\over E-E_n}\right) V_{vpc}\vert n0> 
\ee
Upon using the following representation for the Coulomb propagator \cite{Voloshin}
\be
<{\bf x}\vert {1\over E-H}\vert{\bf y}>=\sum_{l=0}^{\infty}G_{l}( x, y,E)\sum_{m=-l}^{l}
Y_{l}^{m}\left ( {{\bf x}\over x}\right )
{Y_{l}^{\ast}}^{m}\left ( {{\bf y}\over y}\right ) 
\ee
\be
G_{l}( x, y,E)= -4k\mu (2kx)^{l}(2ky)^{l} e^{-k(x+y)}\sum_{n^{\prime}=1}^{\infty}{L_{n^{\prime}-1}^{2l+1}(2kx)
L_{n^{\prime}-1}^{2l+1}(2ky) \Gamma (n^{\prime}) \over (n^{\prime}+l-{\mu \alpha\over k} )
\Gamma (n^{\prime}+2l+1) } \nn
\ee
where $k^2=-2\mu E$, (18) can be split into three pieces
\be
\delta \Psi_{\small n0}({\bf 0})=\delta_{ps} \Psi_{\small n0}({\bf 0})+\delta_{zph} \Psi_{\small n0}({\bf 0})+\delta_{mph} \Psi_{\small n0}({\bf 0})
\ee
The first piece (pole subtraction) corresponds to the term $n^{\prime}=n$ in the sum (20) and it can be calculated using the formulas given above. 
The remaining pieces read 
\bea
\delta_{zph} \Psi_{\small n0}({\bf 0})+\delta_{mph} \Psi_{\small n0}({\bf 0})
 &=& 
 \frac{\alpha}{\pi} \Psi_{\small n0}({\bf 0}) \int_{\small 0}^{\small 1} \, dv \, \frac{v^2 \left ( 1- \frac{v^2}{3} \right )\xi^{n-1}}{(\xi + \sqrt{1-v^2})^{n+1}} \, \sum_{\small n^{\prime}=1,n^{\prime} \neq n}^{\small \infty} \, \frac{n^{\prime}n}{n^{\prime}-n} \nn \\
&& \left ( \frac{\xi}{\xi+ \sqrt{1-v^2}} \right )^{n^{\prime}-1} F \left ( -(n-1),-(n^{\prime}-1);2 ; \frac{{1-v^2}}{\xi^2} \right ).  
\eea
Again the hypergeometric function above reduces to a polynomial. For $n=1$ 
it reduces in fact to 1 and the sum over $n^{\prime}$ can be carried out explicitly. We obtain: 
\bea 
\delta_{zph} \Psi_{\small 10}({\bf 0})+\delta_{mph} \Psi_{\small 10}({\bf 0})
&=& \frac{\alpha}{\pi} \Psi_{\small 10}({\bf 0}) \int_{\small 0}^{\small 1}  dv  \frac{v^2 \left ( 1- \frac{v^2}{3} \right )}{(\xi + \sqrt{1-v^2})^2} \left \{  \frac{\xi}{\sqrt{1-v^2}}+ \log \left (\frac{\xi+\sqrt{1-v^2}}{\sqrt{1-v^2}} \right ) \right \}.  
\eea
where the first term corresponds to the zero photon exchange and the second one to the multiphoton exchange.
Again the change of variable $v\rightarrow \cos\theta$ and a number of manipulations allow us to obtain (6) from the above.

\end{document}